# Exotic dielectric behaviors induced by pseudo-spin texture in magnetic twisted bilayer

Yu-Hao Shen (沈宇皓)[1], Wen-Yi Tong (童文旖)[1], He Hu (胡鹤)[1], Jun-Ding Zheng (郑君鼎)[1],

Chun-Gang Duan (段纯刚)[*1,2]

[1]State Key Laboratory of Precision Spectroscopy and Key Laboratory of Polar Materials and Devices, Ministry of Education, Department of Electronics, East China Normal University, Shanghai, 200241, China

[2]Collaborative Innovation Center of Extreme Optics, Shanxi University, Taiyuan, Shanxi 030006, China

* E-mail: cgduan@clpm.ecnu.edu.cn

**ABSTRACT** Twisted van der Waals bilayers provide an ideal platform to study the electron correlation in solids. Of particular interest is the 30° twisted bilayer honeycomb lattice system, which possesses an incommensurate moiré pattern and uncommon electronic behaviors may appear due to the absence of phase coherence. Such system is extremely sensitive to further twist and many intriguing phenomena will occur. In this work, based on first-principles calculations we show that, for further twist near 30°, there could induce dramatically different dielectric behaviors of electron between left and right twisted cases. Specifically, it is found that the left and right twists show suppressed and amplified dielectric response under vertical electric field, respectively. Further analysis demonstrate that such exotic dielectric property can be attributed to the stacking dependent charge redistribution due to twist, which forms twist-dependent pseudospin textures. We will show that such pseudospin textures are robust under small electric field. As a result, for the right twisted case, there is almost no electric dipole formation exceeding the monolayer thickness when the electric field is applied. Whereas for the left case, the system could even demonstrate negative susceptibility, i.e. the induced polarization is opposite to the applied field, which is very rare in the nature. Such findings not only enrich our understanding on moiré systems but also open an appealing route toward functional 2D materials design for electronic, optical and even energy storage devices.

## INTRODUCTION

Twistronics rooted in the twisted bilayer van der Waals crystals is of both theoretical and technological importance [1-6]. Recent advances in fabrication of atomically thin materials have successfully realized the interlayer twist, giving rise to an alternative way to modulate layered potential. In such twisted 2D systems, moiré pattern with long period [7-13] is introduced by misoriented stacking. The patterned interlayer coupling in van der Waals crystals significantly modifies the low-energy band structure. For instance, it is characterized by the flat bands in twisted bilayer graphene (tBLG) and the Dirac velocity become zero several times with the change of the magic-angle [14-17]. The strong electron correlation in tBLG yields various fascinating physical behaviors, such as the transitions from semimetal to Mott insulator and even unconventional superconductivity [18,19].

Unlike the magic-angle tBLG system with commensurate superlattice, when the twist angle is 30° between the two layers, the singularity appears, making the system lack of long period and become incommensurate with 12-fold rotational symmetry [20-24]. More interestingly, for the bilayer transition-metal dichalcogenides (TMD) 2H-$MX_2$ (M=Mo, W and X=S, Se) system [25-32], due to the existence of different stackings, turning left and right near 30° twist angle corresponds to different atomic structures. Consequently, the two generated moiré systems possess different phase modulations of the electronic wavefunction in the superlattices, rendering stacking dependent properties that are absent in tBLG system [22,24,33]. For instance, the modulation on exchange interaction [34,35] form different patterns in two twisted cases. Further, as magnetism is extremely sensitive to the structural change in two-dimensional systems [34,36,37], we expect the non-spin-degenerate twisted bilayer system with large intrinsic exchange coupling could demonstrate more fascinate electron correlation behaviors. Such considerations therefore motivate us to extend the moiré TMD bilayers to magnetic system.

In this work, based on a typical magnetic system 2H-$VSe_2$, we show that when we apply a vertical electric field to break the potential balance between layers, there will emerge exotic dielectric behaviors for twisted bilayer near 30°. For the right twisted case, there is almost no electric dipole formation exceeding the monolayer thickness when the electric field is applied. Whereas for the left case, the system could even demonstrate negative susceptibility. Further investigations demonstrate that the deep origin of the special dielectric response is that small electric field could hardly change the pseudospin texture in these two partner moiré systems. All above found are based on performing first-principles calculations. The potential applications using such dielectric properties are proposed as well.

## RESULTS

We choose the ferromagnetic 2H-$VSe_2$, a typical 2D ferrovalley material [38,39], as a platform and set its bilayer to be antiferromagnetically coupled, which is the ground state of most pristine bilayer system [40,41]. In this way, such an antiferrovalley bilayer system [40,41] possess both spin-valley locking [42-45] and spin-layer locking [36,46]. Taking 30±2.2° twisted cases for example and the in-plane shift between the two layers is set to be zero. The top view of the incommensurate (infinite) and commensurate (finite) lattice structures are shown in **Figure. 1(a)**, in which the supercell of the two commensurate partners possess the same size (**Figure. 1(b)**). Here we define that the $\delta < 0$ ($> 0$) case corresponds to the right (left) rotation of top layer with respect to bottom layer in the $30+\delta°$ twisted bilayers. Note that the band structures for left and right twisted bilayer systems correspond to different superpositions of the monolayer bands. This can be understood by the mirror symmetry of the two first Brillouin Zone (BZ) of the supercell along **Γ-K** axis [26], illustrated as the grey hexagons in **Figure. 1(b)**. In considering of the bands folding into this first BZ, we obtain that for left twisted supercell, the $K_+$ states is superposed by $K_+$ valley states of bottom layer and $K_-$ valley states of top layer (H-type like stacking), whereas for the right twisted supercell, it is superposed by the two $K_+$ valley states of both two layers (R-type like stacking).

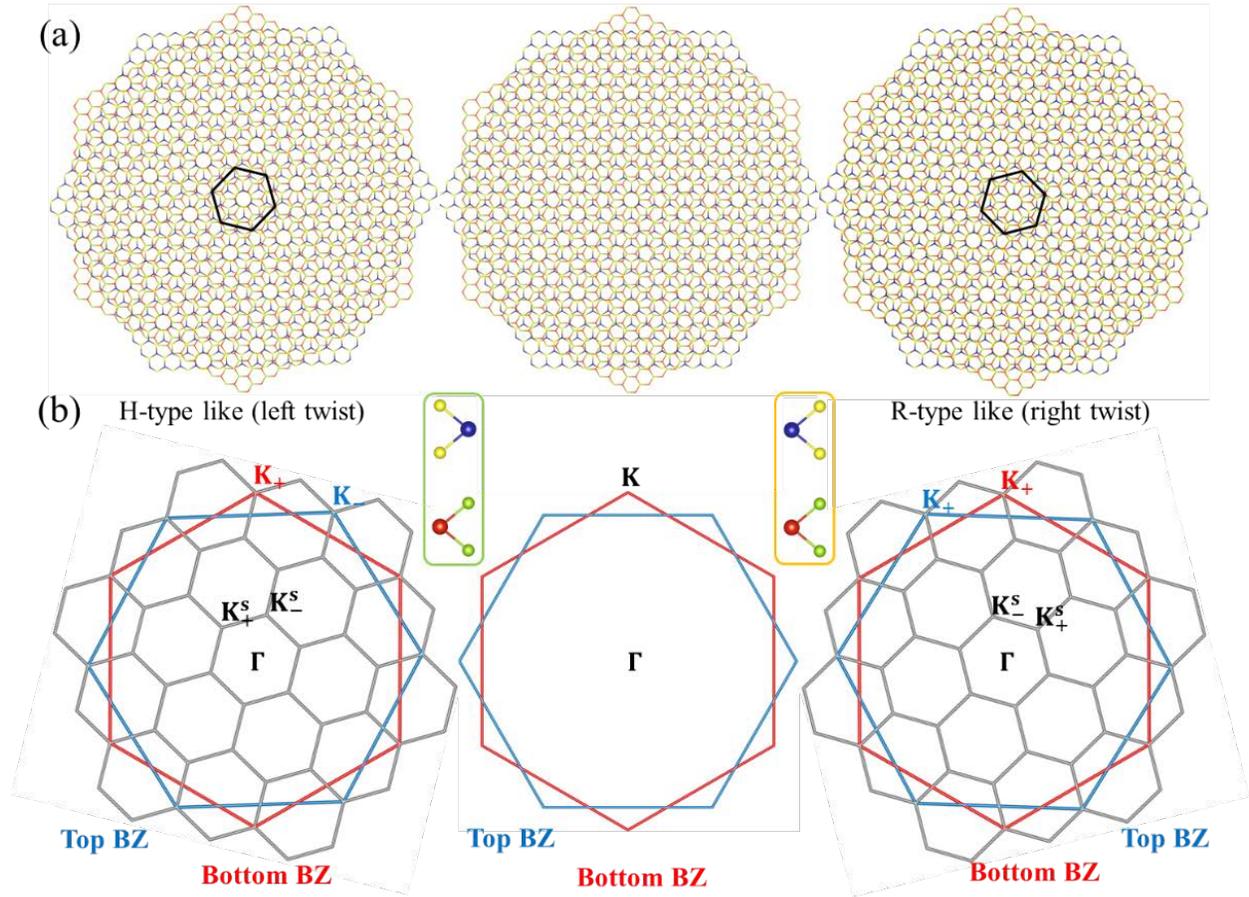

**Figure. 1.** (a) Top view of the bilayer 2H-VSe$_2$ commensurate case of left twist (left panel), right twist (right panel) and incommensurate case of 30° twist case (middle panel). Grass-green and red sites of the honeycomb represent bottom layer selenium and vanadium atoms, respectively. Yellow

and blue sites in the hexagons represent top layer selenium and vanadium atoms, respectively. (b) The first Brillouin Zone (BZ) of the superlattice is shown as the grey hexagons.

For these two cases, we study the dielectric response of the two systems by applying the vertical electric field. Here, the positive field is defined as pointing from top layer to bottom layer. Surprisingly, though the two kind of systems only differ by no more than 5° rotations, we find dramatically different response under external field, illustrated as the significantly different shift of the energy bands under electric field. As shown in **Figure. 2**, when $E = 0.02$ V/Å, there is an about 0.1 eV splitting of the valence band maxima of the two layers at **Γ** for the right twisted case, 10 times larger than that for left twisted case (~ 0.01 eV). It is remarkable that for left twisted case, the bands are insensitive to external electric field, or in other words, it resists to the external electric field. Whereas for the right twisted case, there exists an amplifying effect, i.e., even a small electric field could induce metal-insulator transition, much smaller than that in the H-type bilayer case [47].

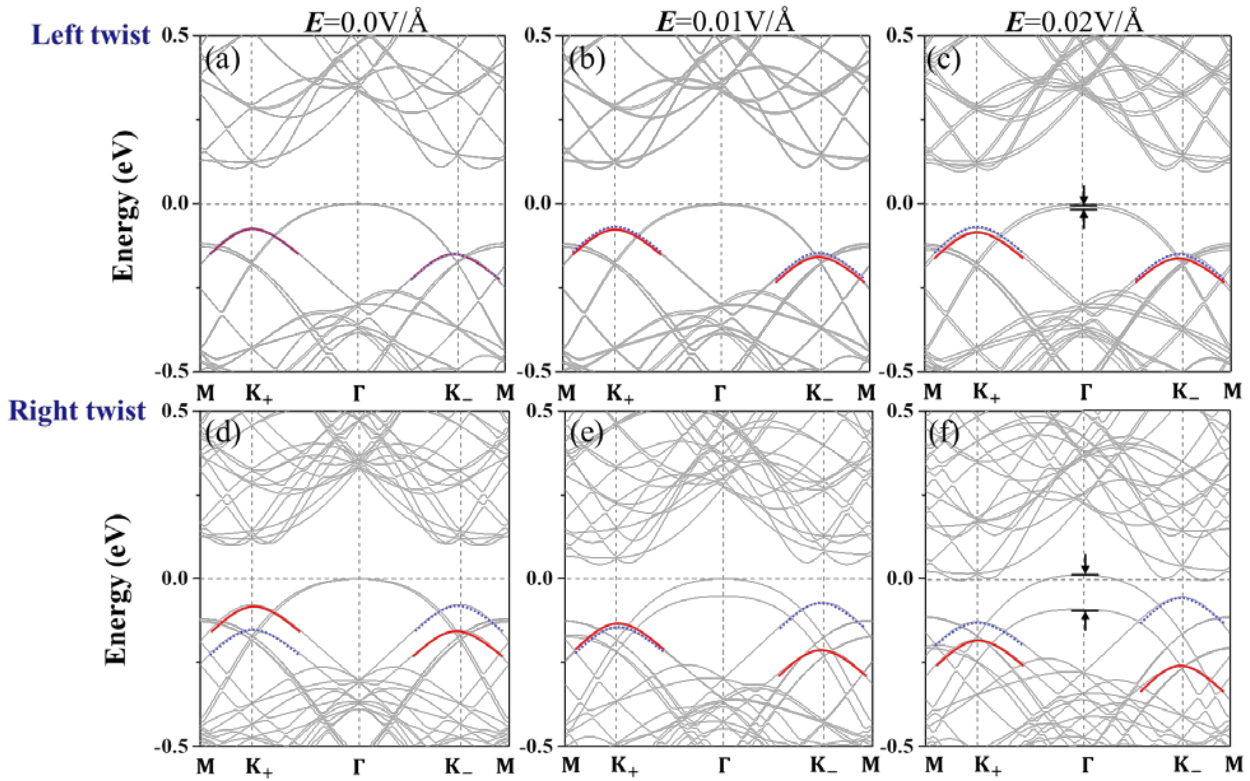

**Figure. 2.** Calculated band structures with spin-orbit coupling of (a-c) 32.2° twist case and (d-f) 27.8° twist case with electric field E applied. (a,d) $E = 0.0$ V/Å. (b,e) $E = 0.01$ V/Å.(e,f) $E = 0.02$ V/Å. For the parabolic description of the **K** valley bands, we use dashed and solid curved lines to denote the bands from top and bottom layer, respectively and the spin-up states (red) and spin down states (blue) is distinguished.

Moreover, the electronic polarization $\Delta P(E)$ obtained by Berry phase method [48] implemented in first-principles calculations is shown in **Figure. 3(a)**. Taking $E = 0.001$ V/Å as an example, the calculated results are $\Delta P = -0.98$ and $\Delta P = 1.26$ for left and right twisted case, respectively. Here, we define $P$ as the dipoles/volume per layer, which is in units of $10^{-4}$ e/ Å$^2$. Compared with the calculated value $\Delta P = 1.03$ of the monolayer system, the twisted bilayers obviously possess suppressed or amplified dielectric polarization. The abnormal dielectric response of the bilayer systems is more clearly seen from the distortion of the electron clouds, as illustrated in the charge difference plot ($\Delta \rho = \rho(E) - \rho(0)$ shown in **Figure. 3(b)**). For left twisted case there induces negative polarization, i.e. the induced dipole is, *opposite* to the applied electric field, in strong contrast to that of the right twisted case, where there is almost no induced dipole that exceeds the monolayer thickness. Indeed, this kind of response is also very strange. As we know, due to the electrostatic effect, the induced dipoles in general are always along the direction of the applied field, for not only the dielectrics but also the magnetic metal film [49,50].

The exotic dielectric difference can be attributed to the initial charge distribution without electric field in such bilayer system, which can be illustrated as the charge distribution difference along $z$ axis between the two twisted cases $\Delta \rho = \rho_R - \rho_L$. Here, $\rho_R$ and $\rho_L$ corresponds to the density distribution projected on $z$ direction in the cell for right and left twisted case, respectively. As shown in **Figure. 3(c)**, there induces net increase of interlayer charge density and net decrease of intralayer charge density when the bilayer system turned from left to right twist. For such antiferromagnetically coupled two ferromagnetic layers, the additional δ-twist from 30° will modulate both the intralayer and interlayer electron cloud distribution, which greatly modulates the magnetic exchange coupling. This patterned modulation is closely related to the magnetic atom sites arrangement of left twisted case (H-type like stacking) or right twisted case (R-type like stacking). And applying vertical electric field tends to induce increase or decrease of the interlayer charge density for left and right twisted cases, respectively.

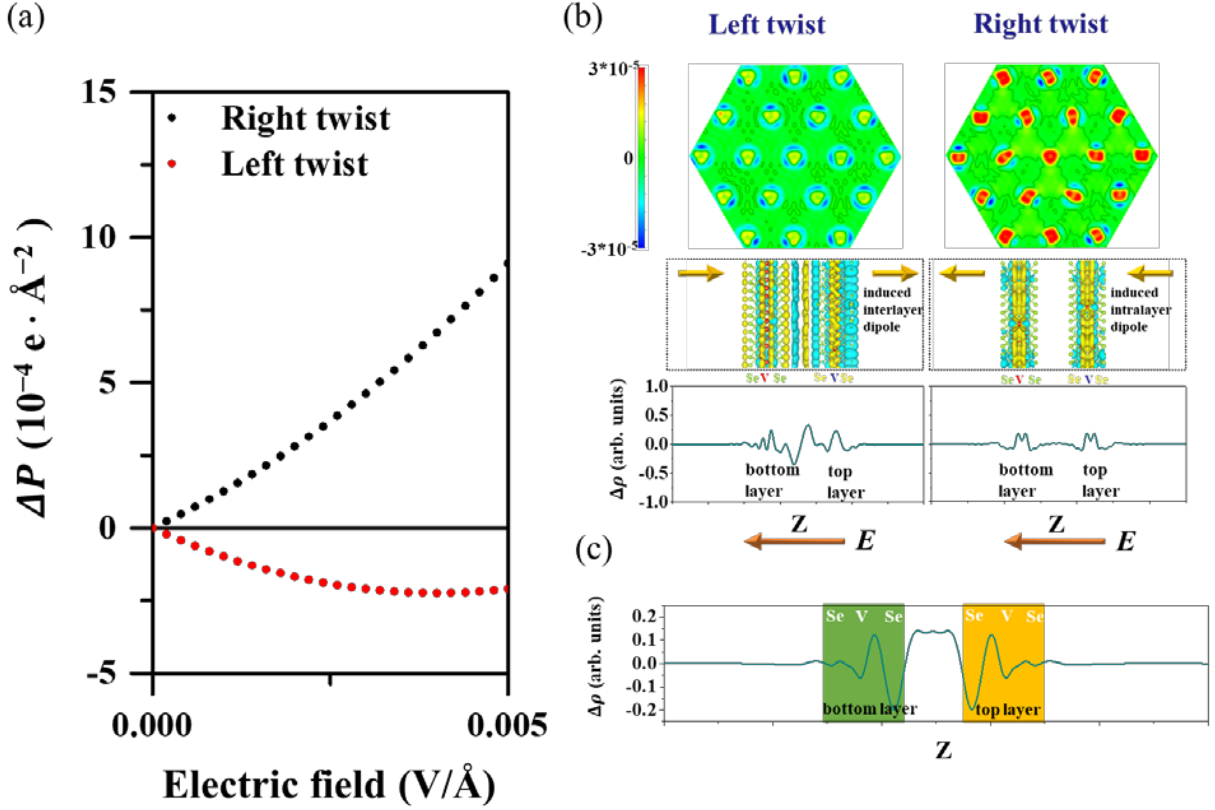

**Figure. 3.** (a) The evolution of calculated $\Delta P(E)$ as a function of electric field $E$. (b)The induced planar charge of the bottom layer under electric field $E = 0.001$ V/Å for left twist case (left) and right twisted case (right). The electric field-induced charge densities $\Delta\rho = \rho(E) - \rho(0)$ in arbitrary units for the two twisted cases. Yellow and cyan colors represent the accumulation and depletion of electrons, respectively. Orange arrows denote the directions of the electric field. (c) Induced average electron charge, $\Delta\rho = \rho_R - \rho_L$ along $z$ axis when the bilayer system turned from right (R) to left (L) twist.

**DISCUSSION**

To understand the underlying mechanism qualitatively, we plot the calculated pseudospin textures of the two cases, which is associated with their orbital moment distribution without applying external field [51]. As shown in **Figure. 4a**, for H-type like stacking there emerges vortex pattern of pseudospin, where the negative ($s_z<0$) and positive ($s_z>0$) one is locked with different spin channel respectively contributed from bottom (spin down) and top layer (spin up). As for R-type like stacking it shows similar vortex structure with different helicity ($\pi/2$ phase difference). Note that when vertical electric field is turned on, only one spin channel will respond this external field because of spin-layer locking effect [36,46]. Then, due to the Rashba spin-orbit coupling (SOC) effect H ($k$, $\sigma$) ~ $k \times E \cdot \sigma$, applying external perpendicular electric field $E$ will create an effective magnetic field $B_{eff}$ which is lying in the plane and normal to the **Γ-K** axis. As can be seen in **Figure. 4b**, this magnetic field tends to generate vortex pseudospin texture like that for R-type like stacking with $\pi/2$ helicity. Therefore, the planar electron density rearranges and it strengthens the intralayer magnetic exchange interaction. The perpendicular change of electron density, especially the part out of the layer, is negligible. That is why there is almost no induced dipole exceeding the monolayer thickness in this case (right panel of **Figure. 3(b)**). In addition, like the electric field effect, the increase of the intralayer magnetic exchange coupling further separates the bands of top and bottom layers. Consequently, the bands split under electric field is doubled, which can be regarded as electric field amplifying effect.

However, for the H-type like stacking, as pointed out above, the applied electric field tends to form vortex pseudospin texture with different helicity, which will destroy the original pseudospin texture. As a consequence of topology protection, the planar electrons now will collectively move out-of-plane to screen or resist the external electric field and therefore *negative* polarization is formed, as clearly shown in the yellow arrows of left panel of **Figure. 3(c)**. Such mechanism keeps the system as close as possible to the unperturbed state ($E = 0$). The negative electric susceptibility for left twisted case is unusual and often appears in metamaterials using artificial design. We assume this novel property may result in useful application in optical waveguide or other optical and electronic devices, e.g., negative capacitor [52]. We find that in either twisted case, the electrostatic energy will be converted in to magnetostatic energy. Therefore, such systems have quite large energy capacity.

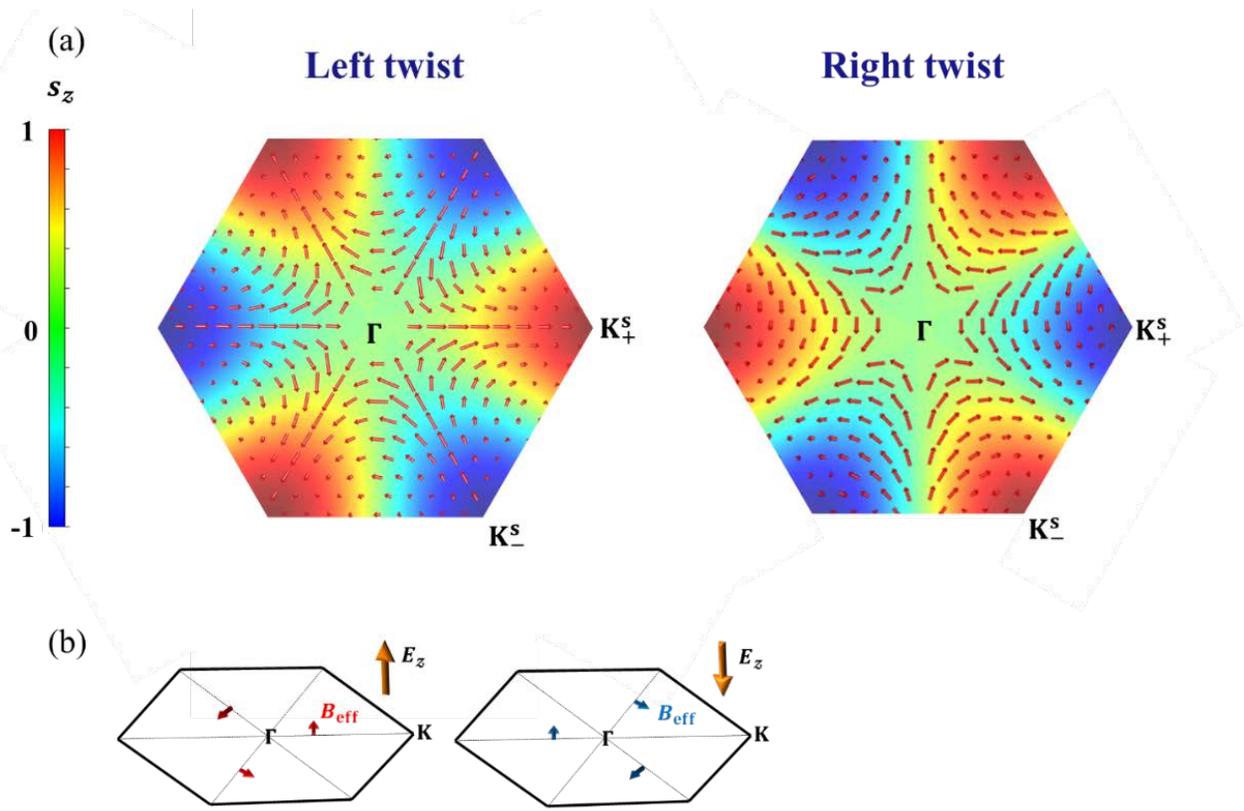

**Figure. 4.** (a) Normalized pseudospin texture in the first Brillouin Zone of the supercell for the highest valence band which contains contributions from bands of both top and bottom layers. The in-plane and out-of-plane components are shown in red arrows and color map, respectively. (b) Schematic plot of Rashba spin-orbit coupling (SOC) effect in our case.

Finally, we should point out that the dielectric properties exhibit almost no difference between the pristine R-type and H-type stacking bilayer systems [41], as no in-plane spin-texture will occur in both systems. When the twist effect is turned on, the dielectric response will change due to the appearance of the pseudo-spin texture. The reason we choose $30 \pm 2.2°$ as representative systems is based on the fact they correspond to commensurate supercells with not too large size and therefore can be handled by our *ab-initio* calculations. Actually this pseudo-spin textures dependent effect can be found in generic twisted system with other angles, but it will be less obvious as twist angle become smaller (δ become larger).

In the second place, though in this study we take 2H-VSe$_2$ bilayer as example to demonstrate the twist effect, it should be ubiquitous for any magnetic bilayers with stacking dependent electronic structures. The magnetic twisted system responds to the external electric field in a magnetic way, which can be regarded as an entirely new magnetoelectric effect, reveals us an intricate pattern how electrostatic and magnetostatic interactions can be entangled when the system become complicated.

To summarize, we show that when an external electric field is applied onto a magnetic twisted bilayer system, there will occur exotic dielectric behaviors of electron depending on the different stackings of two layers, dramatically different from that of monolayer. This unique phenomenon well interprets the philosophy of *More is different* [53]. Based on a platform of 30±2.2° twisted system, we carry out first-principles calculations of band structures, electronic polarization and take analysis on the charge density distribution. There emerges amplification and suppression response to the vertical electric field for right and left twisted case near 30°, respectively. It is demonstrated that this unusual dielectric phenomenon depends on the pseudospin-textures of the twisted systems and there give rise to a novel magnetoelectric effect, i.e., direct electrostatic-magnetostatic energy transforming. We expect more inspiring discoveries on these twisted systems.

**METHODS**

When $\delta = \pm 2.2°$, the corresponding supercell possesses the $\sqrt{13} \times \sqrt{13}$ times size of the unit cell [33]. For the monolayer 2H-VSe$_2$, the intermediate layer of hexagonally arranged vanadium atoms are sandwiched between two atomic layers of selenium. The optimized lattice constant is 3.335Å. From top view, it possesses honeycomb structure and is lack of inversion symmetry. When turned from $\delta = 2.2°$ case to $\delta = -2.2°$ case, for the bilayer structure, the top layer stacked on the bottom layer rotates 180° along the [110] of its supercell. Then we obtain two different stacking bilayers, where they possess twist angle 32.2° and 27.8°.

The *ab-initio* calculations of bilayer VSe$_2$ are performed within density functional theory (DFT) using the projector augmented wave (PAW) method implemented in the Vienna *ab-initio* Simulation Package (VASP) [54]. The exchange-correlation potential is treated in Perdew–Burke–Ernzerhof form[55] of the generalized gradient approximation (GGA-PBE) with a kinetic-energy cutoff of 400 eV. A well-converged 9×9×1 Monkhorst-Pack *k*-point mesh is chosen in self-consistent calculations. The convergence criterion for the electronic energy is $10^{-5}$ eV and the structures are relaxed until the Hellmann–Feynman forces on each atom are less than 1 meV/Å. In our calculations, the SOC effect was explicitly included in the calculations and the dispersion corrected DFT-D2 method[56] is adopted to describe the van der Waals interactions. The external electric field is introduced by planar dipole layer method.

**ACKNOWLEDGMENT**

This work was supported by the National Key Research and Development Program of China (2017YFA0303403), Shanghai Science and Technology Innovation Action Plan (No. 19JC1416700), the NSF of China 11774092, ECNU Multifunctional Platform for Innovation.

# REFERENCES

[1] C. Berger, et al. 2006 *Science* **312** 1191
[2] Z. H. Ni, Y. Y. Wang, T. Yu, Y. M. You, and Z. X. Shen 2008 *Phys. Rev. B* **77** 235403
[3] Z. Yan, Z. W. Peng, Z. Z. Sun, J. Yao, Y. Zhu, Z. Liu, P. M. Ajayan, and J. M. Tour 2011 *Acs Nano* **5** 8187
[4] L. M. Xie, H. L. Wang, C. H. Jin, X. R. Wang, L. Y. Jiao, K. Suenaga, and H. J. Dai 2011 *J. Am. Chem. Soc.* **133** 10394
[5] R. Q. Zhao, Y. F. Zhang, T. Gao, Y. B. Gao, N. Liu, L. Fu, and Z. F. Liu 2011 *Nano. Res.* **4** 712
[6] S. Carr, D. Massatt, S. Fang, P. Cazeaux, M. Luskin, and E. Kaxiras 2017 *Phys. Rev. B* **95** 075420
[7] J. Kang, J. Li, S. S. Li, J. B. Xia, and L. W. Wang 2013 *Nano. Lett.* **13** 5485
[8] G. H. Li, A. Luican, J. M. B. L. dos Santos, A. H. Castro Neto, A. Reina, J. Kong, and E. Y. Andrei 2010 *Nat. Phys.* **6** 109
[9] M. M. Ugeda, et al. 2014 *Nat. Mater.* **13** 1091
[10] R. Bistritzer and A. H. MacDonald 2011 *P. Natl. Acad. Sci. USA* **108** 12233
[11] M. Yankowitz, J. M. Xue, D. Cormode, J. D. Sanchez-Yamagishi, K. Watanabe, T. Taniguchi, P. Jarillo-Herrero, P. Jacquod, and B. J. LeRoy 2012 *Nat. Phys.* **8** 382
[12] J. M. Xue, J. Sanchez-Yamagishi, D. Bulmash, P. Jacquod, A. Deshpande, K. Watanabe, T. Taniguchi, P. Jarillo-Herrero, and B. J. Leroy 2011 *Nat. Mater.* **10** 282
[13] A. Luican, G. H. Li, A. Reina, J. Kong, R. R. Nair, K. S. Novoselov, A. K. Geim, and E. Y. Andrei 2011 *Phys. Rev. Lett.* **106** 126802
[14] G. TramblydeLaissardiere, D. Mayou, and L. Magaud 2012 *Phys. Rev. B* **86** 125413
[15] Y. Cao, et al. 2018 *Nature* **556** 80
[16] B. Padhi, C. Setty, and P. W. Phillips 2018 *Nano. Lett.* **18** 6175
[17] T. M. R. Wolf, J. L. Lado, G. Blatter, and O. Zilberberg 2019 *Phys. Rev. Lett.* **123** 096802
[18] K. Kim, et al. 2017 *P. Natl. Acad. Sci. USA* **114** 3364
[19] Y. Cao, V. Fatemi, S. Fang, K. Watanabe, T. Taniguchi, E. Kaxiras, and P. Jarillo-Herrero 2018 *Nature* **556** 43
[20] C. R. Woods, et al. 2014 *Nat. Phys.* **10** 451
[21] A. V. Lebedev, I. V. Lebedeva, A. M. Popov, and A. A. Knizhnik 2017 *Phys. Rev. B* **96** 085432
[22] W. Yao, et al. 2018 *P. Natl. Acad. Sci. USA* **115** 6928
[23] I. V. Lebedeva, A. V. Lebedev, A. M. Popov, and A. A. Knizhnik 2016 *Phys. Rev. B* **93** 235414
[24] S. J. Ahn, et al. 2018 *Science* **361** 782
[25] K. H. Liu, et al. 2014 *Nat. Commun.* **5** 4966
[26] M. L. Lin, et al. 2018 *Acs Nano* **12** 8770
[27] F. C. Wu, T. Lovorn, and A. H. MacDonald 2017 *Phys. Rev. Lett.* **118** 147401
[28] B. X. Cao and T. S. Li 2015 *J. Phys. Chem. C* **119** 1247
[29] H. Y. Yu, G. B. Liu, J. J. Tang, X. D. Xu, and W. Yao 2017 *Sci. Adv.* **3** e1701696
[30] F. C. Wu, T. Lovorn, and A. H. MacDonald 2018 *Phys. Rev. B* **97** 035306
[31] M. H. Naik and M. Jain 2018 *Phys. Rev. Lett.* **121** 266401
[32] Y. Wang, Z. Wang, W. Yao, G. B. Liu, and H. Y. Yu 2017 *Phys. Rev. B* **95** 115429
[33] P. Moon and M. Koshino 2013 *Phys. Rev. B* **87** 205404
[34] N. Sivadas, S. Okamoto, X. D. Xu, C. J. Fennie, and D. Xiao 2018 *Nano. Lett.* **18** 7658
[35] P. Jiang, C. Wang, D. Chen, Z. Zhong, Z. Yuan, Z. Y. Lu, and W. Ji 2018 *arXiv:1806.09274*
[36] B. Huang, et al. 2018 *Nat. Nanotechnol.* **13** 544
[37] J. R. Schaibley, H. Y. Yu, G. Clark, P. Rivera, J. S. Ross, K. L. Seyler, W. Yao, and X. D. Xu 2016 *Nat. Rev. Mater.* **1** 16055
[38] W. Y. Tong, S. J. Gong, X. G. Wan, and C. G. Duan 2016 *Nat. Commun.* **7** 13612
[39] X. W. Shen, W. Y. Tong, S. J. Gong, and C. G. Duan 2018 *2d Mater.* **5** 011001
[40] M. Esters, R. G. Hennig, and D. C. Johnson 2017 *Phys. Rev. B* **96** 235147
[41] W. Y. Tong and C. G. Duan 2017 *npj Quantum Mater.* **2** 47
[42] W. Yao, D. Xiao, and Q. Niu 2008 *Phys. Rev. B* **77** 235406
[43] D. Xiao, G. B. Liu, W. X. Feng, X. D. Xu, and W. Yao 2012 *Phys. Rev. Lett.* **108** 196802
[44] D. Xiao, W. Yao, and Q. Niu 2007 *Phys. Rev. Lett.* **99** 236809
[45] X. D. Xu, W. Yao, D. Xiao, and T. F. Heinz 2014 *Nat. Phys.* **10** 343
[46] A. M. Jones, H. Y. Yu, J. S. Ross, P. Klement, N. J. Ghimire, J. Q. Yan, D. G. Mandrus, W. Yao, and X. D. Xu 2014 *Nat. Phys.* **10** 130

[47] S. J. Gong, C. Gong, Y. Y. Sun, W. Y. Tong, C. G. Duan, J. H. Chu, and X. Zhang 2018 *P. Natl. Acad. Sci. USA* **115** 8511
[48] R. D. King-Smith and D. Vanderbilt 1993 *Phys. Rev. B* **47** 1651
[49] C. G. Duan, J. P. Velev, R. F. Sabirianov, Z. Q. Zhu, J. H. Chu, S. S. Jaswal, and E. Y. Tsymbal 2008 *Phys. Rev. Lett.* **101** 137201
[50] Y. H. Shen, Y. X. Song, W. Y. Tong, X. W. Shen, S. J. Gong, and C. G. Duan 2018 *Adv. Theory Simul.* **1** 1800048
[51] X. W. Zhang, Q. H. Liu, J. W. Luo, A. J. Freeman, and A. Zunger 2014 *Nat. Phys.* **10** 387
[52] S. Salahuddin and S. Dattat 2008 *Nano. Lett.* **8** 405
[53] P. W. Anderson 1972 *Science* **177** 393
[54] G. Kresse and J. Furthmuller 1996 *Comp. Mater. Sci.* **6** 15
[55] J. P. Perdew, K. Burke, and M. Ernzerhof 1996 *Phys. Rev. Lett.* **77** 3865
[56] S. Grimme 2006 *J. Comput. Chem.* **27** 1787